\documentstyle[12pt,world_sci]{article}

\newcommand{\beqn}{\begin{eqnarray}}
\newcommand{\eeqn}{\end{eqnarray}}
\def\spose#1{\hbox to 0pt{#1\hss}}
\def\lsim{\mathrel{\spose{\lower 3pt\hbox{$\mathchar"218$}}
     \raise 2.0pt\hbox{$\mathchar"13C$}}}
\def\gsim{\mathrel{\spose{\lower 3pt\hbox{$\mathchar"218$}}
     \raise 2.0pt\hbox{$\mathchar"13E$}}}
\def\simpropto{\mathrel{\spose{\lower 3pt\hbox{$\mathchar"218$}}
     \raise 2.0pt\hbox{$\propto$}}}
\def\mpl{M_{\rm P}}

\def\beq{\begin{equation}}
\def\eeq{\end{equation}}
\def\barr{\begin{array}}
\def\earr{\end{array}}
\def\and{\qquad {\rm and } \qquad}
\def\etal{ {\it et al.} }

\def\eg{ {\it e.g.} }
\def\ZPC#1#2#3{{\sl Z.~Phys.} {\bf C#1}, #2 (#3)}
\def\PTP#1#2#3{{\sl Prog. Theor. Phys.} {\bf #1}, #2 (#3)}
\def\PRL#1#2#3{{\sl Phys. Rev. Lett.} {\bf #1}, #2 (#3)}
\def\PRD#1#2#3{{\sl Phys. Rev.} {\bf D#1}, #2 (#3)}
\def\PLB#1#2#3{{\sl Phys. Lett.} {\bf B#1}, #2 (#3)}
\def\PREP#1#2#3{{\sl Phys. Rep.} {\bf #1}, #2 (#3)}
\def\NPB#1#2#3{{\sl Nucl. Phys.} {\bf B#1}, #2 (#3)}

\def\tev{{\rm TeV }}
\def\gev{{\rm GeV }}

\def\fb{{\rm fb}^{-1}}
\def\vev#1{{\langle#1\rangle}}
\def\msusy{M_{\rm SUSY}}
\def\mgut{M_{\rm GUT}}

\def\mz{m_{\rm z}}
\def\mw{m_{\rm w}}
\def\mhl{m_{h^0}}
\def\mha{m_{A^0}}

\def\barp{\overline{p}}

\def\tanb{\tan\beta}

\catcode`\@=11
\long\def\@makefntext#1{
\protect\noindent \hbox to 3.2pt {\hskip-.9pt
$^{{\ninerm\@thefnmark}}$\hfil}#1\hfill}                

\def\@makefnmark{\hbox to 0pt{$^{\@thefnmark}$\hss}}  

\def\ps@myheadings{\let\@mkboth\@gobbletwo
\def\@oddhead{\hbox{}
\rightmark\hfil\ninerm\thepage}
\def\@oddfoot{}\def\@evenhead{\ninerm\thepage\hfil
\leftmark\hbox{}}\def\@evenfoot{}
\def\sectionmark##1{}\def\subsectionmark##1{}}

\renewcommand{\thefootnote}{\fnsymbol{footnote}}

\newcounter{sectionc}\newcounter{subsectionc}\newcounter{subsubsectionc}
\renewcommand{\section}[1] {\vspace*{0.6cm}\addtocounter{sectionc}{1}
\setcounter{subsectionc}{0}\setcounter{subsubsectionc}{0}\noindent
        {\normalsize\bf\thesectionc. #1}\par\vspace*{0.4cm}}
\renewcommand{\subsection}[1] {\vspace*{0.6cm}\addtocounter{subsectionc}{1}
        \setcounter{subsubsectionc}{0}\noindent
        {\normalsize\it\thesectionc.\thesubsectionc. #1}\par\vspace*{0.4cm}}
\renewcommand{\subsubsection}[1]
{\vspace*{0.6cm}\addtocounter{subsubsectionc}{1}
        \noindent
{\normalsize\rm\thesectionc.\thesubsectionc.\thesubsubsectionc.
        #1}\par\vspace*{0.4cm}}

\newcounter{appendixc}
\newcounter{subappendixc}[appendixc]
\newcounter{subsubappendixc}[subappendixc]

\renewcommand{\appendix}[1] {\vspace*{0.6cm}
        \refstepcounter{appendixc}
        \setcounter{figure}{0}
        \setcounter{table}{0}
        \setcounter{equation}{0}
        \renewcommand{\thefigure}{\Alph{appendixc}.\arabic{figure}}
        \renewcommand{\thetable}{\Alph{appendixc}.\arabic{table}}
        \renewcommand{\theappendixc}{\Alph{appendixc}}
        \renewcommand{\theequation}{\Alph{appendixc}.\arabic{equation}}
        \noindent{\bf Appendix \theappendixc #1}\par\vspace*{0.4cm}}

\def\abstracts#1{{

\centering{\begin{minipage}{12.2truecm}\footnotesize\baselineskip=12pt\noindent
        \centerline{\footnotesize ABSTRACT}\vspace*{0.3cm}
        \parindent=0pt #1
        \end{minipage}}\par}}

\renewenvironment{thebibliography}[1]
        {\begin{list}{\arabic{enumi}.}
        {\usecounter{enumi}\setlength{\parsep}{0pt}
\setlength{\leftmargin 1.25cm}{\rightmargin 0pt}
         \setlength{\itemsep}{0pt} \settowidth
        {\labelwidth}{#1.}\sloppy}}{\end{list}}

\newcounter{itemlistc}
\newcounter{romanlistc}
\newcounter{alphlistc}
\newcounter{arabiclistc}

\newcommand{\fcaption}[1]{
        \refstepcounter{figure}
        \setbox\@tempboxa = \hbox{\footnotesize Fig.~\thefigure. #1}
        \ifdim \wd\@tempboxa > 6in
           {\begin{center}
        \parbox{6in}{\footnotesize\baselineskip=12pt Fig.~\thefigure. #1}
            \end{center}}
        \else
             {\begin{center}
             {\footnotesize Fig.~\thefigure. #1}
              \end{center}}
        \fi}

\newcommand{\tcaption}[1]{
        \refstepcounter{table}
        \setbox\@tempboxa = \hbox{\footnotesize Table~\thetable. #1}
        \ifdim \wd\@tempboxa > 6in
           {\begin{center}
        \parbox{6in}{\footnotesize\baselineskip=12pt Table~\thetable. #1}
            \end{center}}
        \else
             {\begin{center}
             {\footnotesize Table~\thetable. #1}
              \end{center}}
        \fi}

\def\@citex[#1]#2{\if@filesw\immediate\write\@auxout
        {\string\citation{#2}}\fi
\def\@citea{}\@cite{\@for\@citeb:=#2\do
        {\@citea\def\@citea{,}\@ifundefined
        {b@\@citeb}{{\bf ?}\@warning
        {Citation `\@citeb' on page \thepage \space undefined}}
        {\csname b@\@citeb\endcsname}}}{#1}}

\newif\if@cghi
\def\cite{\@cghitrue\@ifnextchar [{\@tempswatrue
        \@citex}{\@tempswafalse\@citex[]}}
\def\citelow{\@cghifalse\@ifnextchar [{\@tempswatrue
        \@citex}{\@tempswafalse\@citex[]}}
\def\@cite#1#2{{$\null^{#1}$\if@tempswa\typeout
        {IJCGA warning: optional citation argument
        ignored: `#2'} \fi}}

 1
 1
 1

\font\ninerm=cmr9

\begin{document}
\begin{flushright}
hep-ph/9506408\\
MPI-PhT/95-56\\
June 1995
\end{flushright}

\centerline{\normalsize\bf SUPERSYMMETRY AT PRESENT AND FUTURE
COLLIDERS\footnote{
to be published in the Proceedings of the Ringberg Workshop:
  ``Dark Matter in the Universe'', Mar. 6-10, 1995}}

\centerline{\footnotesize Ralf Hempfling}
\baselineskip=13pt
\centerline{\footnotesize\it Max-Planck-Institut f\"ur Physik,
Werner-Heisenberg-Institut,}
\baselineskip=12pt
\centerline{\footnotesize\it F\"ohringer Ring 6, 80805 Munich, Germany}
\centerline{\footnotesize E-mail: hempf@iws180.mppmu.mpg.de}
\vspace*{0.3cm}

\vspace*{0.9cm}
\abstracts{
The theoretical expectations for the supersymmetric particle spectrum
is reviewed and a brief overview on present constraints on
supersymmetric models from collider experiments is presented.
Finally, we discuss the discovery potential of future colliders experiments.
}

\normalsize\baselineskip=15pt
\setcounter{footnote}{0}
\renewcommand{\thefootnote}{\alph{footnote}}
\section{Introduction}

The standard model of elementary particle physics (SM) is in
excellent agreement with present experimental results.
Nonetheless, the theory suffers from a variety of
theoretical shortcomings and is generally believed to
be the low energy effective theory
describing the physics at and below the scale of spontaneous
electro-weak symmetry breaking given by
the mass of the $Z$ boson, $\mz = 91.187~\gev$\cite{pdg}.
In particular, finding an explanation for
the hierarchy between $\mz$ and the Planck scale,
$\mpl = 10^{19}~\gev$ is considered a sever problem
and most theorists expect that the mechanism to solve this
problem should
manifest itself already at energies below $1~\tev$.
Presently, the most popular and the most promising candidate for
this new physics is supersymmetry (SUSY)~\cite{susyreview}.

In supersymmetric theories the quadratically divergent
higher order contributions to scalar mass terms
from bosons (fermions) are automatically
canceled by the contributions of the fermionic (bosonic)
superpartner.

This implies immediately
that the number of fields has to be doubled with respect to
the SM.
In addition, two Higgs doublets are required
in order to give masses to down-type and up-type fermions
and to cancel the Higgsino (= fermionic superpartner of the Higgs
bosons) contributions to the triangle anomalies.
The particle content
of the minimal supersymmetric extension of the SM (MSSM)
together with their quantum numbers
is presented in table~\ref{pcontent}.

\begin{table}[t]
\tcaption{The MSSM particle content.}\label{pcontent}
$$
\begin{tabular}{lcccccc}
\hline
\hline
\multicolumn{2}{c}{Superfields}
  & SM fields & Superpartner & SU(3)$_c$ & SU(2)$_L$ & U(1)$_Y$\\
\hline
\multicolumn{2}{c}{\underline{Gauge Multiplets}} & {} & {} & {} &{} & {}\\
\multicolumn{2}{c}{$\widehat G$} & $g$ & $\tilde g$ & 8 & 1 & 0\\
\multicolumn{2}{c}{$\widehat W$} & $W$ & $\widetilde W$ & 1 & 3 & 0\\
\multicolumn{2}{c}{$\widehat B$} & $B$ & $\widetilde B$ & 1 & 1 & 0\\
\hline
\multicolumn{2}{c}{\underline{Matter Multiplets}} & {} & {} & {} &{} & {}\\
 {} & $\widehat Q$ & $(u,d)_L$ & $\widetilde Q = (\tilde u, \tilde d)_L$
& $3       $ & $2$ & $\phantom{-}1/3$\\
quarks & $\widehat U^c$ & $\bar u_L$ & $\widetilde U^c = \tilde u_R^*$
 & ${\bar 3}$ & $1$ & $-4/3$\\
{} & $\widehat D^c$ & $\bar d_L$ & $\widetilde D^c = \tilde d_R^*$
& ${\bar 3}$ & $1$ & $\phantom{-}2/3$\\
\multicolumn{1}{c}{\rule[0mm]{0mm}{6mm} {}}
& $\widehat L$ & $(\nu, e)_L$ & $\widetilde L = (\tilde \nu, \tilde e)_L$
& $1       $ & $2$ & $-1$\\
\raisebox{1.5ex}[-1.5ex]{leptons}&
$\widehat E^c$ & $\bar e_L$ & $\widetilde E^c = \tilde e_R^*$
 & 1 & $1$ & $\phantom{-}2$\\
\multicolumn{1}{c}{\rule[0mm]{0mm}{6mm} {}}
& $\widehat H_1$ & $H_1$ & $\left(\widetilde H^0_1,\widetilde H^-_1\right)$
& $1 $ & $2$ & $-1$\\
\raisebox{1.5ex}[-1.5ex]{Higgs}
& $\widehat H_2$ & $H_2$ & $\left(\widetilde H^+_2,\widetilde H^0_2\right)$
 & 1 & $2$ & $\phantom{-}1$\\
\hline
\end{tabular}
$$
\end{table}

Furthermore, all the couplings of the superpartners
are determined by SUSY and in particular one finds that
they have to be mass degenerate with their SM counterpart.
This later requirement is in clear contradiction
with the measurement of the $Z$ line-shape at
LEP which tells us that there
are no new particles with masses below about $\mz/2$
except maybe for electro-weak singlets such as
the fermionic partners of the photon and of the gluon\cite{pdg}.

In general it is very difficult to construct realistic
low energy models where SUSY is broken spontaneously.
Instead it became standard to break SUSY
explicitly via soft SUSY breaking terms\cite{soft-susy}.
These terms are thought to arises from
spontaneous SUSY breaking via dilaton or moduli fields (Superstring
Theories) or via gravitational coupling
to a "hidden sector", where the spontaneous  SUSY breaking
takes place without posing any experimental constraints
on our visible world.

With these soft SUSY breaking terms we can give masses
to all the superpartners and thus evade all experimental
constraints. In particular, we can give a negative
squared mass to the Higgs bosons needed for the spontaneous
electro-weak symmetry breaking (EWSSB).

The Higgs boson is the last missing building block of the
SM\cite{berndi}
and its discovery is not a priori an indication for
SUSY\footnote{%
However, the discovery of a fundamental scalar would severely limit
any alternative solution to the hierarchy problem such as
compositeness models or techi-color models.}.
However, there are fundamental differences
between the SM and the MSSM Higgs sector that might allow us
to disentangle the two models.
As stated above, the Higgs sector of the MSSM consists of two
Higgs doublets which after EWSSB reduces to
two CP-even Higgs bosons, $h^0$ and $H^0$,
one CP-odd Higgs boson, $A^0$, and a pair of
charged Higgs boson, $H^\pm$.
Because of SUSY the Higgs potential contains only two independent
parameters typically chosen to be $m_{A^0}$ and
the ratio of the Higgs vacuum expectation values, $\tanb \equiv
\vev{H_2}/\vev{H_1}$.
It also predicts a well defined upper limit for
the lightest MSSM Higgs boson\cite{mhl}
\beqn
m_{h^0} \lsim \mz +\hbox{radiative corrections}
\simeq 130~\gev\,,
\eeqn
such that non-discovery of a Higgs boson below about 130$~\gev$
would rule out the MSSM.
This bound can be evaded by coupling the Higgs fields to
an additional gauge singlet.
In this extended model the Higgs mass becomes a free parameter
depending on this coupling.
However, an upper limit
analogous to the triviality bound in the SM remains
if one requires this coupling to remain perturbative up to some high
scale (say, $\mpl$)\cite{nmssm}.

The soft SUSY breaking terms can be written as follows\cite{hlw}
\beqn
V_{soft} = V_{0} + V_{1/2} + V_{A} + V_{B}\,,
\label{soft}
\eeqn
where $V_0 = \sum_\phi m_{i j}^2 \phi^\dagger_i \phi_j$
gives a mass to
all the spin 0 particles (the sum is over all Higgs bosons,
squarks and sleptons),
$V_{1/2} = \sum_i m_i \psi_i \psi_i + h.c.$
gives a mass to all the spin 1/2 partners of the gauge bosons
($\psi_i = \widetilde B$, $\widetilde W$ and $\tilde g$),
$V_{A}$ describes the trilinear Higgs-squark-squark interactions,
and $V_{B}$ describes the Higgs mixing term which is usually replaced in
favor of $\tanb$.
With these additional terms we can construct a model that satisfies all
experimental constraints.

Unfortunately, the predictability
of these models is very limited due to the large number of free
parameters. The situation can be improved by making
assumptions about the origin of these parameters.
In minimal supergravity models one usually assumes
that the soft mass terms generated via gravity are universal
at $\mpl$ for all fields with the same spin
and that the squark-squark-Higgs interactions are
proportional to the corresponding quark-quark-Higgs Yukawa interactions.
That means that the SUSY particle spectrum is determined
in terms of only four parameters
$m_0^2$, $m_{1/2}$, $A$, corresponding to the four terms in
eq.~\ref{soft}\footnote{the Higgs mass parameter of the
superpotential, $\mu$, is determined by fixing $\mz$.}.
This universality of the soft SUSY breaking terms
is broken via renormalization group
evolution from $\mpl$ to $\mz$ and one again obtains a
non-degenerate particle spectrum. However, this
evolution is predictable if we assume that
the MSSM is the full theory all the way to
$\mpl$ without any intermediate scales\footnote{%
Clearly this is not the case in grand unified theories (GUTs)
but the hope was that the numerical results will not be much affected
by the RG integration between $\mpl$ and $\mgut = 2\times 10^{16}~\gev$.
This assumption may be an oversimplification as was pointed out recently
in refs.~\citenum{non-univ}.}.
We obtain
\beqn
m_{\tilde F}^2 = C_0^{F} m_0^2 + C_{1/2}^{F} m_{1/2}^2 + C_D^F D\,,
\label{coeffs}
\eeqn
where we have neglected the $A$ parameter which is mainly important
for the left/right mixing of the top squarks. Here,
we have defined $D = \mw^2 \cos 2\beta$
and $C_D^F = T_3 -\tan^2\theta_{\rm w} Y_F/2$ ($Y_F$ denotes the
U(1)$_Y$ hypercharges shown in
table~\ref{pcontent} and $\theta_{\rm w}$ is
the Weinberg angle).
In the case of small Yukawa couplings
these coefficients are given by $C_0^F \simeq 1$ and
$C_{1/2}^F \simeq 6.5$, 6, 6, 0.5, 0.15,
for $F = Q, U, D, L, E$.
The coefficients for the third generation $F = Q_3, U_3$
(and possibly $F = D_3$ in the limit of large $\tanb$)
can be reduced by up to about 50\%
through the effects of a large top Yukawa coupling.
It is this effect of the top Yukawa coupling that
renders the coefficients $C_0^{H_2}$ and $C_{1/2}^{H_2}$
negative and, hence, destabilizes the symmetric
minimum. As a result, $H_2$ (and via mixing also $H_1$)
acquire a non-vanishing VEV that break the electro-weak
symmetry spontaneously\cite{radssb}
The fact, that over a large region of the parameter space only
the Higgs fields and non of the scalar superpartners acquire a VEV
is one of the reasons of for the popularity of this scenario.

Another reason is the fact that the gauge couplings
meet within the experimental errors at a single scale\cite{amal},
$\mgut \simeq 2 \times 10^{16}~\gev$
which allows to embed the SM gauge-groups
into a single unified gauge group such as $SU(5)$\cite{su5}.
In SUSY-GUT models\cite{gutreview} the universality of
the gaugino masses in $V_{1/2}$ is not only a (model-depending)
assumption but an inevitable consequence of gauge invariance.
A violation of the resulting low energy prediction
\beqn
{M_{\tilde g}\over m_{1/2}}:
{M\over m_{1/2}}:
{M^\prime\over m_{1/2}}
= {\alpha_s\over \alpha_{GUT}}:
{\alpha_{em}\over \sin^2\theta_{\rm w}\alpha_{GUT}}:
{5 \alpha_{em}\over 3\cos^2\theta_{\rm w}\alpha_{GUT}}
\simeq 2.5: 0.8: 0.4\,,
\eeqn
would be a indication against
minimal SUSY-GUT.
Note that the gauginos of a broken gauge symmetry are no
mass eigenstates but they obtain a SUSY invariant mass
via mixing with the Higgsinos.
The resulting neutral (charged) mass eigenstates are the
neutralinos (charginos).
In the SUSY limit (ie: $M^\prime = M = \mu = 0$ and
$\tanb = 1$)
one linear combination of $\widetilde B$ and $\widetilde W^3$ combines with
one linear combination of $\widetilde H_1^0$ and $\widetilde H_2^0$
to form a Dirac fermion with mass $\mz$ while the other
fields remain massless.
On the other hand, in the limit of a large SUSY
breaking scale $\msusy \gg \mz$ the
the mass eigenstates are the gauginos with majorana masses $M^\prime$, $M$
and the Higgsinos with a Dirac mass $\mu$.
The chargino sector is analogous with
$\mz$ replaced by $\mw$, etc.

An important property of the MSSM is its minimality not
only with respect to its particle content but also
with respect to the allowed particle interactions.
In general, any model constrained solely by gauge invariance will
contain new interactions that violate
lepton number and baryon number. These interactions
have to be suppressed below the present limits
by imposing an additional symmetry.
The most popular candidate for such a symmetry,
$R$-parity\cite{susyreview}, is defined such that all
superpartners change sign while the SM particles remain
invariant. $R$ parity conservation
implies that the lightest supersymmetric partner
(LSP) is stable and, hence, will contribute to the dark matter (DM)
of the universe\cite{stefan}.

The existence of a stable LSP is not only important
for DM search but also for the SUSY search at colliders.
Most experimental analysis assumes that the superpartner
decay (possibly via cascade decays) into the LSP which escapes from
the detector undetected. The evidence for such an event
would be the missing transverse energy that had been carried away by the
LSP.

\newpage

\section{SUSY in Present and Future Collider Experiments}

\begin{table}[t]
\tcaption{Present and Future Colliders.}\label{tabcollider}
$$
\begin{tabular}{lcrrl}
\hline
\hline
name & Type & $\sqrt{s}$ & $\int {\cal L} d t$ & date\\
\hline
\multicolumn{5}{l}{\underline{$e^+e^-$ Colliders:}} \\
LEP-I  & circular & $\mz     $           & $0.1~\fb$           & now\\
LEP-II & circular & $180~\gev$           & $0.5~\fb$           & 1996/7\\
NLC    & linear   & $0.5 \sim 2~\tev$ & $50\sim 200~\fb$ & 2005/10\\
\hline
\multicolumn{5}{l}{\underline{Hadron Colliders:}} \\
Tevatron (CDF and D0)    & $p \barp$ & $1.8~\tev$ & $0.1~\fb$  & now \\
Tevatron (main injector) & $p \barp$ & $2~\tev$  & $1~\fb$  & 1999 \\
TeV$^*$                  & $p \barp$ & $2~\tev$  & $10~\fb$ & 2000/1 \\
Di-Tevatron              & $p \barp$ & $4~\tev$  & $20~\fb$ & 2000/1 \\
LHC                      & $p p$    & $14~\tev$  & $10\sim 100~\fb$
& 2004/8\\
\hline
\end{tabular}
$$
\end{table}

Now after we have introduced the theoretical framework
we will present limits on the supersymmetric parameter space
from collider experiments.
The main constraints on the SUSY particle masses come
from high energy $e^+e^-$ or hadron collider experiments.
In addition, there are various other experiments\cite{otherx}
whose primary importance for SUSY is to constrain individual
(e.g. $R$ parity violating) interactions rather than
particle masses.
In table~\ref{tabcollider}\
we have listed all relevant high energy collider projects that
are presently running (LEP-I, Tevatron),
approved (LEP-II, main injector, LHC), considered (TeV$^*$, Di-Tevatron)
and under discussion
(NLC with the option for $e^-\gamma$, $e^-e^-$,
and $\gamma \gamma$ collision).

\subsection{SUSY at $e^+e^-$ Colliders}

The clearest constraints come from LEP experiments at CERN
that has collected $O(10^7)$ on-shell $Z$'s.
The high statistics not only to rules out
any squarks and charged sleptons with a mass
below $\mz/2$ but allows also to establish similar lower limits
on the sneutralino masses from the $Z$ line-shape
and some parameter dependent lower limit on the lightest neutralino mass.

LEP-II is expected to start operating
by the end of 1996
and the members of the various working groups
are asked to make discovery and exclusion plots
for a center-of-mass energies of
$\sqrt{s} = 175$, 192, and 205~\gev\cite{lep2}.
Any $e^+e^-$ collider with $\sqrt{s} \gsim 200~\gev$
would have to be linear due to excessive power-loss
of a circular collider via
sychrotron radiation. The discovery potential of
the next linear $e^+e^-$ collider (NLC) with an initial
energy of $\sqrt{s} \gsim 500~\gev$ and a option for an upgrade to
$1.5\sim 2~\tev$ is under continuing study\cite{nlc1}.

The discovery potential of any of these $e^+e^-$
machines for a charged particle is roughly given
by $\sqrt{s}/2$ minus a few \% due to phase-space
suppression near threshold.

The discovery of the lightest Higgs boson is possible for masses
$\mhl \lsim \sqrt{s} - 100~\gev$. This means that non-discovery of a
Higgs boson at LEP-II would constrain the MSSM parameter space
but one would have to wait for the NLC to
rule out the MSSM.

The disadvantage is that squarks can only be produced via the
s-chanal which is suppressed for high energies
and that gluinos can only be produced in squark decay
if $M_{\tilde g} < m_{\tilde q}$.

\subsection{SUSY at Hadron Colliders}

The best constraint on the gluino mass, $M_{\tilde g} > 144~\gev$
(or $M_{\tilde g}>212$ if $M_{\tilde g} = m_{\tilde q}$),
come from the D0 experiment at TEVATRON\cite{gq@pp}

The SUSY discovery potential of was studied by the SUSY working group
for the LHC workshop\cite{lhc}
and more recently in ref.~\citenum{tata}.
The disadvantage is that a Higgs boson with a mass $\mhl \lsim 2 \mw$
decays predominantly into $b\bar b$ pairs and discovery
at a hadron collider in this case is difficult.
In particular, there is a hole in the $\tanb$--$\mha$ plane
for $4\lsim \tanb\lsim 15$ and $100~\gev \lsim \mha \lsim 200~\gev$
where no MSSM Higgs boson can be detected\cite{kz}

\section{Summary}

$e^+e^-$ colliders provide the ideal environment for
the discovery of Higgs bosons and for a spectroscopy of
the electro-weak superpartners (= sleptons, charginos and neutralinos).
The lightest chargino might well be within the range of LEP-II.
At the NLC with $\sqrt{s} = 500~\gev$ one will for sure discover
at least one Higgs boson or rule out the MSSM and any other SUSY-GUT model
that requires perturbativity of the couplings up to $\mgut$.
Many if not all electro-weak superpartners are expected to
lie within the reach of the NLC upgrade with $\sqrt{s} = 1.5\sim 2~\tev$.

On the other hand, the search for colored superpartners
is more promising in hadron colliders because
(a) squarks and gluinos are expected to be heavier (see, eq.~\ref{coeffs})
(b) the production rate at high energies is dominated by t-chanal
exchange of gluinos (squark production) or
of squarks (gluino production) absent in $e^+e^-$ colliders.
Thus, we conclude that $e^+e^-$ and hadron colliders
are very much complimentary in the search for SUSY.

%
%


\newpage

%

\end{document}